\documentclass[aip,preprint]{revtex4-1}
\usepackage{graphicx}
\usepackage{amsfonts}
\usepackage{amsmath}
\usepackage{color}
\usepackage{lineno}
\draft 
\begin{document}
\title{Griffiths phase-like behaviour and spin-phonon coupling in double perovskite Tb$_{2}$NiMnO$_{6}$} 
\author{Harikrishnan S. Nair}
\affiliation{J\"{u}lich Center for Neutron Sciences-2/Peter Gr\"{u}nberg Institute-4, Forschungszentrum J\"{u}lich GmbH, 52425 J\"{u}lich, Germany}
\email{h.nair@fz-juelich.de, krishnair1@gmail.com}
\author{Diptikanta Swain}
\affiliation{Chemistry and Physics of Materials Unit, Jawaharlal Nehru Centre for Advanced Scientific Research, Bangalore 560064, India}
\author{Hariharan N.}
\affiliation{Department of Physics, C.V. Raman Avenue, Indian Institute of Science, Bangalore 560012, India}
\author{Shilpa Adiga}
\affiliation{J\"{u}lich Center for Neutron Sciences-2/Peter Gr\"{u}nberg Institute-4, Forschungszentrum J\"{u}lich GmbH, 52425 J\"{u}lich, Germany}
\author{Chandrabhas Narayana}
\affiliation{Chemistry and Physics of Materials Unit, Jawaharlal Nehru Centre for Advanced Scientific Research, Bangalore 560064, India}
\author{Suja Elizabeth}
\affiliation{Department of Physics, C.V. Raman Avenue, Indian Institute of Science, Bangalore 560012, India}
\date{\today}
\begin{abstract}
The Griffiths phase-like features and the spin-phonon coupling effects observed
in Tb$_2$NiMnO$_6$ are reported.
The double perovskite compound crystallizes in monoclinic $P2_1/n$
space group and exhibits a magnetic phase transition at $T_c \sim$ 111~K
as an abrupt change in magnetization.
A negative deviation from ideal Curie-Weiss law exhibited
by 1/$\chi(T)$ curves and less-than-unity susceptibility exponents
from the power-law analysis of inverse susceptibility are reminiscent of Griffiths phase-like features.
%
Arrott plots derived from magnetization isotherms
support the inhomogeneous nature of magnetism in this material.
%
The observed effects originate from antiferromagnetic interactions
which arise from inherent disorder in the system.
Raman scattering experiments display no magnetic-order-induced phonon
renormalization below $T_c$ in Tb$_2$NiMnO$_6$ which is different
from the results observed in other double perovskites and is correlated
to the smaller size of the rare earth.
The temperature evolution of full-width-at-half-maximum for the {\it stretching} mode at 645 cm$^{-1}$
presents an anomaly which coincides with the magnetic transition temperature
and signals a close connection between magnetism and lattice in this material.
\end{abstract}
\maketitle 
\section{Introduction}
Double perovskites $R_2BB'$O$_6$ ($R$ = rare earth; $B,B'$ = transition metal)
are interesting systems owing to the variety of
phenomena they display including large magnetocapacitance,
magnetoresistance, cationic ordering, high temperature structural phase transitions
and the predicted multiferroic properties.
\cite{rogadoam_17_2225_2005,bulljpcm_15_4927_2003,das_prl_100_186402_2008,singhprl_100_087601_2008,kumar_prb_82_134429_2010}
%
%
One of the most studied double perovskite, La$_2$NiMnO$_6$ displays large,
magnetic-field-induced changes in resistivity and dielectric constant at 280~K --
a temperature much higher than previously reported for such couplings.
\cite{rogadoam_17_2225_2005}
The combination of multiple functionalities with magnetic, dielectric and lattice
degrees of freedom makes double perovskites a material of current interest
with device application potential.
\cite{hashisaka_jmmm_310_1975_2007,guo_apl_89_022509_2006}
$R_2BB'$O$_6$ compounds are reported to crystallize either in
monoclinic $P2_1/n$ space group, in which case, layers of
$B^{2+}$ and $B'^{4+}$ alternate periodically; or in orthorhombic
$Pbnm$ structure where $B^{3+}$ and $B'^{3+}$ are randomly distributed in the lattice.
\cite{singhapl_91_012503_2007,bulljpcm_15_4927_2003}
Most of the ordered double perovskites exhibit ferromagnetism (FM) which
originates from the superexchange interaction between ordered $B^{2+}$ and
$B'^{4+}$ ions.
\cite{goodenoughpr_100_564_1955,dassprb_68_064415_2003,rogadoam_17_2225_2005}
In ordered La$_2$NiMnO$_6$, the Ni$^{2+}$ ($t^5_{2g} e^4_g$)--O--Mn$^{4+}$ ($t^3_{2g}e^0_g$)
magnetic exchange leads to prominent
ferromagnetic interactions.
However, even in the ordered state, certain percentage of $B$ and $B'$
cations interchange their respective crystallographic
positions leading to what is known as {\it antisite disorder} in these materials.
Monte Carlo simulations as well as experimental investigations
have revealed that this disorder markedly influences the magnetic properties
and can lead to secondary magnetic phases at low temperature.
\cite{ogaleapl_75_537_1999,zhouapl_96_262507_2010,singhapl_91_012503_2007}
%
%
Antisite disorder can introduce additional antiferromagnetic exchange interactions in the form
of Ni$^{2+}$--O--Ni$^{2+}$ or Mn$^{4+}$--O--Mn$^{4+}$ which can result in reduction
of ferromagnetic saturation magnetization, and such interactions,
which get modulated by cationic size mismatch at the La site have been
observed through magnetic measurements on La$_2$NiMnO$_6$.
\cite{zhouapl_96_262507_2010}
It must be noted that short range ferromagnetic correlations above
$T_c$ have been observed in La$_2$NiMnO$_6$,
confirmed by the anomalous softening of phonon modes due to spin-phonon
coupling which extended up to high temperatures.
\cite{iliev_apl_90_151914_2007,guoprb_79_172402_2009,zhouapl_91_172505_2007}
X ray magnetic circular dichroism studies revealed clear signals above
$T_c$ in La$_2$NiMnO$_6$ indicating the presence of short-range FM correlations.
\cite{guoprb_79_172402_2009}
Majority of the existing literature on double perovskites addresses La-based
systems
whereas the effects of smaller, heavier magnetic rare earth at the $R$ site
have received less attention.
Recently investigations on a series of $R_2$NiMnO$_6$
compounds found that the Ni--O--Mn bond length and bond angle,
which are directly involved in superexchange interactions,
are significantly modified with reduction in $R$-size.
\cite{boothmrb_44_1559_2009}
As a consequence, multiple magnetic interactions can develop in the system and
lead to inhomogeneous magnetic states similar to the clustered or phase separated
states observed in perovskite manganites.
%
To strike a comparison with the case of manganites,
quenched disorder arising from the random distribution
of cations of different sizes and charges
introduces phase inhomogeneity through size mismatch of cations and
bond disorder of the Mn-O-Mn network, there by, paving the way for Griffiths phase.
\cite{salamonprb_68_014411_2003,salamonprl_88_197203_2002}
Many manganites exhibit Griffiths phase-like feature, for example,
(La$_{1-y}$Pr$_y$)$_{0.7}$Ca$_{0.3}$Mn$^{16/18}$O$_3$,
\cite{jiangepl_84_47009_2008}
Sm$_{0.5}$Sr$_{0.5}$MnO$_{3}$ nano-manganites,
\cite{zhoujpc_115_1535_2011}
and Pr$_{0.5}$Sr$_{0.5}$Mn$_{1-y}$Ga$_y$O$_3$.
\cite{pramanikprb_81_024431_2010}
The cationic disorder arising from the mixed occupancy at the $B$
site and the smaller size of the rare earth could lead to similar
effects in double perovskites.
\\
%
%
The phonon spectra of double perovskites have been
actively investigated using Raman spectroscopy methods
for example, in La$_2$NiMnO$_6$ and La$_2$CoMnO$_6$ bulk and thin films.
\cite{guo_prb_77_174423_2008,bulljssc_177_2323_2004,guo_apl_89_022509_2006,ilievprb_75_104118_2007}
Strong spin-phonon coupling has been reported in single crystals
and epitaxial thin films of La$_2$NiMnO$_6$.
\cite{ilievjap_106_023515_2009,iliev_apl_90_151914_2007}
%
%
The effect of cationic size of the rare earth on the phonon properties
has not been investigated yet.
In a recent Raman study on Pr$_2$NiMnO$_6$, Truong {\it et al.} observed
that with a smaller rare earth than La at the $R$ site, the spin-phonon
coupling is weakened and the physical properties of rare earth double
perovskite were a function of the type of $R$.
\cite{truongjpcm_23_052202_2011} 
%
%
Recent theoretical prediction
\cite{kumar_prb_82_134429_2010}
about the multiferroicity in Y$_2$NiMnO$_6$
also points towards the importance of the type of $R$ in double perovskites.
It was found theoretically that changing $R$ from La to Y drives the ground state from ferromagnetic
to antiferromagnetic $E*$ type which breaks inversion symmetry
generating electric polarization.
It is, therefore, rewarding to investigate the magnetic and phonon properties of double
perovskites upon change in ionic size of $R$.
In this report we present the results of detailed magnetization and Raman scattering
experiments on Tb$_2$NiMnO$_6$ to explore the effect of substituting a
magnetic rare earth at the $R$ site on the magnetism and spin-phonon coupling.
\section{Experimental Section}
Polycrystalline Tb$_2$NiMnO$_6$ used in the present study was
prepared by conventional solid state synthesis.
%
The precursors Tb$_{2}$O$_{3}$, NiO and MnO$_{2}$ (3N purity or higher)
weighed in stoichiometric amounts were mixed and ground together
in a mortar and was heat treated initially at 1000$^{\circ}$C for 24~h
and further at 1300$^{\circ}$C for 36~h, after re-grinding.
Powder X ray diffractogram (PXD) was obtained by a Philips X'Pert diffractometer
(Cu $K_{\alpha}$) and was analyzed by Rietveld method
\cite{rietveld}
using FULLPROF suite of programs.
\cite{carvajal}
Temperature dependent dc magnetization measurements were performed in
a commercial SQUID magnetometer (Quantum Design) in zero field-cooled (ZFC)
and field-cooled (FC) cycles with applied fields of 20, 100, 500, 20~kOe and 40~kOe.
In addition, at 20~Oe, field-cooled warming/cooling (FCW/FCC) curves
were also recorded.
Raman scattering experiments were performed in a custom-built Raman spectrometer
\cite{pavancs_93_778_2007}
using a 532~nm frequency-doubled solid state Nd-YAG laser with 8~mW power.
The measurements were carried out in back scattering geometry in the
temperature range 77 to 300~K.
\section{Results and Discussion}
\begin{figure}[!h]
\centering
\includegraphics[scale=0.62]{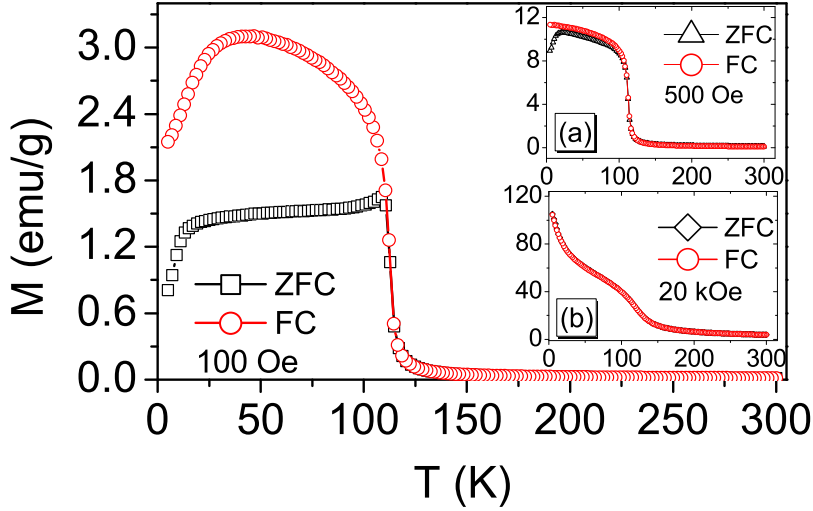}
\caption{Main panel: The ZFC and FC magnetization curves of Tb$_2$NiMnO$_6$ at 100~Oe.
The magnetic transition is evident at $T_c \sim$ 111~K.
The insets (a) and (b) show the ZFC and FC curves at 500 and 20~kOe
respectively.}
\label{fig_mag}
\end{figure}

\begin{figure}[!h]
\centering
\includegraphics[scale=0.40]{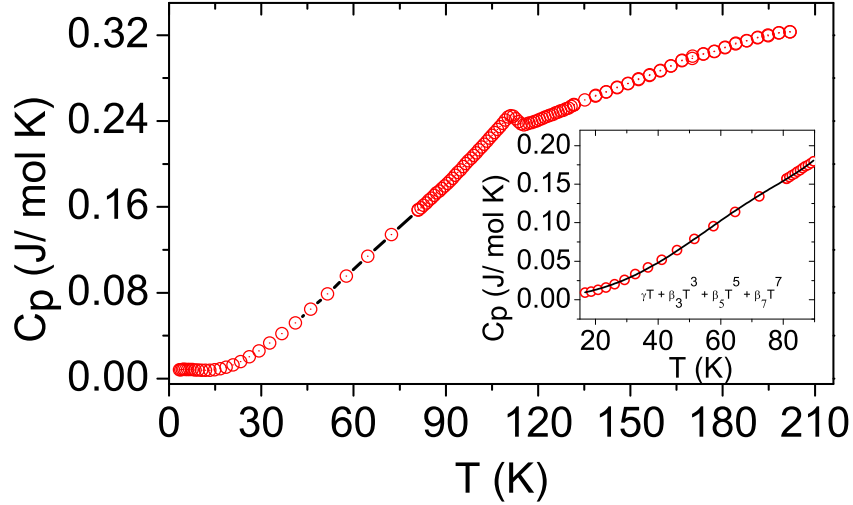}
\caption{Main panel: The specific heat of Tb$_2$NiMnO$_6$ confirming
the phase transition of the Ni/Mn lattice at $T_c \sim$ 111~K.
A broad hump-like feature is observed at low temperature which is related to Schottky effect.
The inset represents the fit to the low temperature $C_p$ assuming the model
described in the text.}
\label{fig_cp}
\end{figure}

\begin{figure}[!h]
\centering
\includegraphics[scale=0.35]{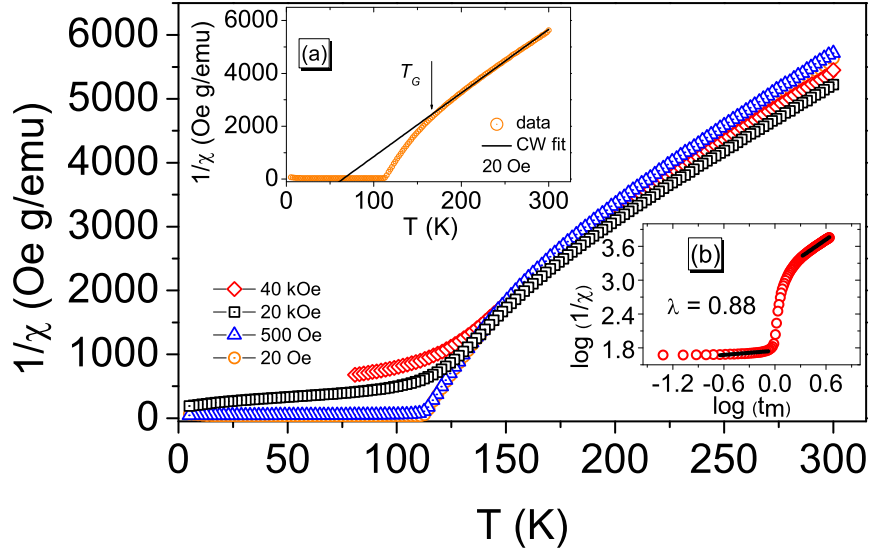}
\caption{Main panel: The inverse susceptibility 1/$\chi(T)$ at different applied
fields 20~Oe, 500~Oe, 20~kOe and 40~kOe. As the applied magnetic field increases,
the downward deviation is suppressed. The inset (a) presents the
CW analysis of 1/$\chi(T)$ at 20~Oe holds only at $T \gg T_c$.
Inset (b) displays log-log plot of the power law analysis
of low-field magnetic susceptibility, $\chi^{-1}(T) \propto (T - T^{R}_C)^{1-\lambda}$. The
black solid lines represent the fit to the experimental data following the power law.}
\label{fig_chi}
\end{figure}

\begin{figure}[!h]
\centering
\includegraphics[scale=0.60]{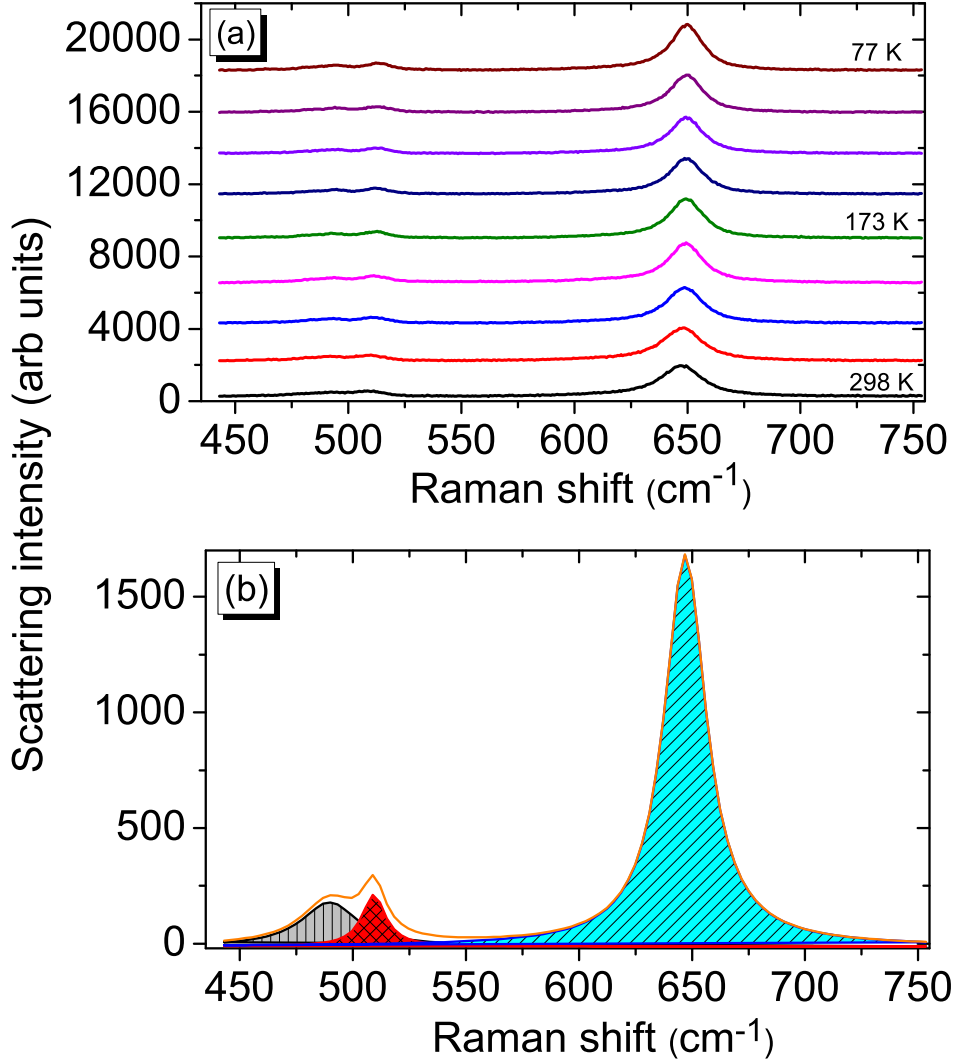}
\caption{(a) Observed Raman scattering intensity curves of Tb$_2$NiMnO$_6$
at different temperatures. (b) The intensity at 298~K as a function of Raman shift,
displaying peaks at $\sim$650~cm$^{-1}$, 510~cm$^{-1}$ and 490~cm$^{-1}$ (thick line).
The shaded areas represent a curve-fit to the observed spectrum using Lorenzian line shape.}
\label{fig_raman}
\end{figure}

\begin{figure}[!h]
\centering
\includegraphics[scale=0.60]{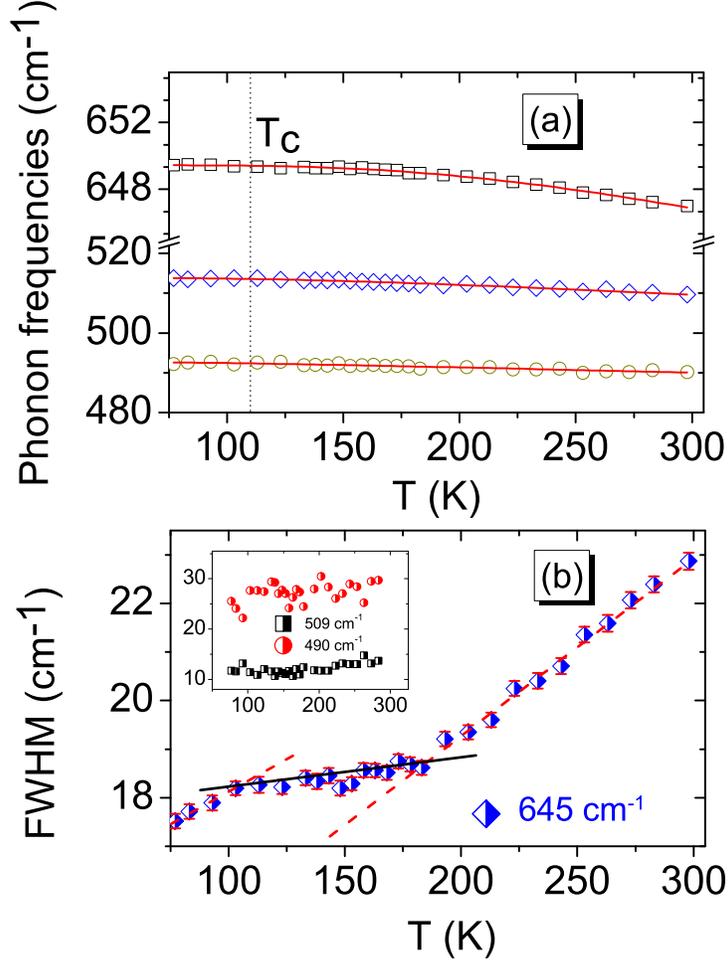}
\caption{(a) The temperature variation of phonon frequencies in Tb$_2$NiMnO$_6$. The red line
is the fit according to the anharmonic phonon-phonon scattering.
A vertical black dotted line marks the position of $T_c\cong$111~K.
(b) The FWHM for the peak at 645 cm$^{-1}$ calculated from the experimental Raman intensity as a function of
temperature. Note the three different slopes in three different temperature ranges. The red dotted lines
show typical behaviour of FWHM while the black solid line represents the anomalous region. The inset
displays curves that correspond to peaks at 490 and 509 cm$^{-1}$ which do not show
a slope change.}
\label{fig_fwhm}
\end{figure}

The crystal structure of Tb$_2$NiMnO$_6$ (TNMO) was refined in monoclinic space
group $P2_1/n$ (space group no.: 14) with an agreement factor,
$\chi^2$ = 3.74.
The refined lattice parameters obtained from the analysis,
$a$ = 5.2699(2)~$\AA$, $b$ = 5.5425(7)~$\AA$ and $c$ = 7.5251(2)~$\AA$
and the monoclinic angle $\beta$ = 89.79$^\circ$ are in reasonable agreement
with the previous report
\cite{boothmrb_44_1559_2009}.
\begin{figure}[!h]
\centering
\includegraphics[scale=0.35]{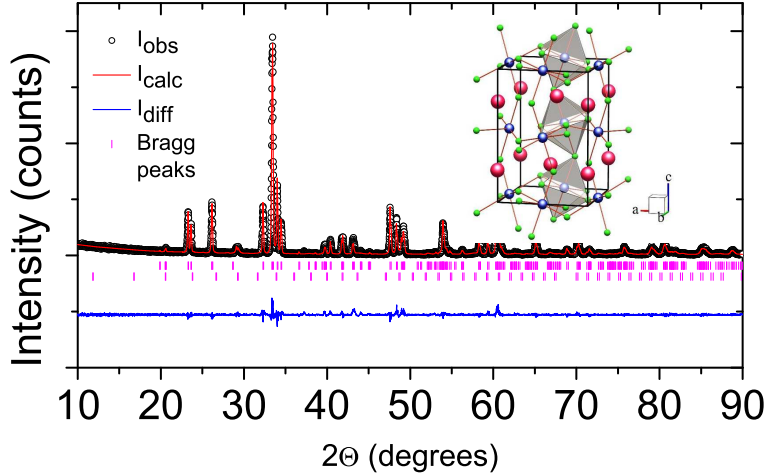}
\caption{The experimental x ray diffraction pattern of Tb$_2$NiMnO$_6$ along with
the results of Rietveld analysis. A minor impurity phase of Tb$_2$O$_3$ was identified
and quantified as 4 wt$\%$. The inset presents a schematic of the crystal structure
in monoclinic $P21/n$ space group. The blue spheres represent Ni/Mn; the green are oxygen
and the pink are Tb. The network of corner-sharing octahedra is indicated.}
\label{fig_str}
\end{figure}
The observed x ray diffraction pattern is analysed using Rietveld method
\cite{rietveld}
in FULLPROF  code
\cite{carvajal}.
The results are presented in Fig. \ref{fig_str} along with a structural
diagram depicting the schematic of the crystal structure.
A minor impurity phase of Tb$_2$O$_3$ (4 wt$\%$) was identified
in the x ray pattern.
The structural parameters and selected bond distances and angles from
the analysis are presented in Table \ref{tbl_str}.
The crystal structure adopted by $R_2BB'$O$_6$ compounds
depends on cationic size, charge and the $R/B$ radius ratio
which are empirically quantified in tolerance factor.
A monoclinic unit cell is favoured if the tolerance factor,
$t = \frac{\frac{r_R + r_{R'}}{2} + r_o}{\sqrt{2}(\frac{r_B + r_{B'}}{2} + r_o)}$
(where $r_R, r_B$ and $r_o$ are ionic radii of rare earth, transition metal
and oxygen, respectively), is less than unity.
With a $t$ = 0.858, a monoclinic unit cell is empirically expected for TNMO
(the ionic radii values for calculating $t$ were taken from
\cite{shannon}).
The bond distances and angles obtained from the structural
refinement compiled in Table \ref{tbl_str} are comparable to the
reported values of Ni$^{2+}$--O and Mn$^{4+}$--O distances
\cite{boothmrb_44_1559_2009}
there by indicating a nearly-ordered Ni$^{2+}$/Mn$^{4+}$ arrangement.
A quantitative estimate of the valence states of the cations
was obtained by calculating the bond valence sums (BVS).
The calculated values, presented in Table \ref{tbl_str} compare well with
those reported
\cite{boothmrb_44_1559_2009}
for the Mn$^{4+}$-- O and Ni$^{2+}$-- O bonds
thereby supporting a predominant Ni$^{2+}$/Mn$^{4+}$
cationic arrangement for TNMO. \\
The $M(ZFC)$ and $M(FC)$ magnetization curves of Tb$_2$NiMnO$_6$
at applied field of 100~Oe are presented in the main panel of
Fig. \ref{fig_mag} while the insets (a) and (b) show the same curves
at higher applied fields of 500 and 20~kOe respectively.
The $ZFC$ and $FC$ arms in Fig. \ref{fig_mag} show a split at the
magnetic transition temperature, $T_{c}\cong$ 111~K where the system enters
magnetically ordered state.
On increasing the applied magnetic field, the $ZFC/FC$ arms begin to merge and at 20~kOe
complete merging is attained.
With the application of magnetic field the
magnetization at low temperature is enhanced,
typical of systems with a ferromagnetic component.
At low applied fields, the $M(ZFC)$ shows a downturn
below $\approx$ 15~K which decreases as the applied field
increases to 20~kOe.
Other double perovskites like Nd$_2$CoMnO$_6$ display
such a downturn which originates from the
anti-parallel alignment of Tb moment with respect to the Ni/Mn moments.
\cite{sazonovjpcm_19_046218_2007}
\\
%
\begin{table}[!h]
\centering
\caption{The structural parameters and selected bond distances and bond angles of Tb$_2$NiMnO$_6$ at room temperature.
The atomic positions were Tb $4e$(x,y,z),  Ni $2c$(0.5,0,0.5), Mn $2d$(0.5,0,0) and O $4e$(x,y,z).
$TM$ stands for the transition metal at the $B$ site, Co or Mn. BVS indicates bond valence sum values.}
\vspace{2mm}
\begin{tabular}{|l|l|} \hline\hline
Space group        & $P2_1/n$                             \\
Space group number & 14                                   \\
Lattice parameters & $a$ = 5.2699(2)~$\AA$, $b$ = 5.5425(7)~$\AA$, $c$ = 7.5251(2)~$\AA$, $\beta$ = 89.79$^\circ$  \\
Bond distance      & TM-O $\approx$ 1.99 \AA    \\
Bond angle         & TM-O1-TM  $\approx$ 140 $^\circ$  \\
BVS                & Tb = 3.6, Ni = 2.4, Mn = 4.1                       \\ \hline\hline
\end{tabular}
\label{tbl_str}
\end{table}
The transition temperature observed in
magnetization measurements was confirmed
through specific heat, $C_p(T)$ as presented in Fig. \ref{fig_cp},
where a peak was observed at 111~K.
Apart from the peak at $T_c$ and a low temperature hump that
arises from the Schottky effect, the $C_p$ displays no anomalies
or peaks from any impurities.
The analysis of low temperature $C_p$ was performed by assuming
a polynomial expansion,
$C_p$ = $\gamma$T + $\beta_3$T$^3$ + $\beta_5$T$^5$ + $\beta_7$T$^7$.
Our fitting gave $\gamma$ = 0.42  mJ/mol K$^2$
for the linear term and $\beta_3$ = 6.2 $\times$ 10$^{-7}$ J/mol K$^4$
for the lattice term.
The $\gamma$ value observed in Tb$_2$NiMnO$_6$ is smaller than the value
observed in La$_{0.4}$Ca$_{0.6}$MnO$_3$ manganite showing Griffiths phase-like
properties.
\cite{lu_jap_103_07F714_2008}
The low value for $\gamma$ also suggests a low contribution from
electronic conduction in this insulating double perovskite.
%
%
%
%
\\
In the main panel of Fig. \ref{fig_chi} the 1/$\chi(T)$ plots at different
applied magnetic fields of 20~Oe, 500~Oe, 20~kOe
and 40~kOe derived from dc magnetization measurements are presented.
The downturn observed in the 1/$\chi(T)$ curves is reminiscent of systems described by Griffiths phase (GP),
\cite{salamonprl_88_197203_2002,salamonprb_68_014411_2003}
which was originally proposed for randomly diluted Ising ferromagnets.
\cite{griffithsprl_23_17_1969}
In the original formulation, Griffiths assumed randomly diluted Ising ferromagnets
with nearest-neighbour exchange bonds of strength $J$ and 0
distributed with a probability $p$ and $1-p$ respectively.
Long range FM order is established only above a percolation threshold ($p > p_c$) in a
reduced $T_C$ which is below the ordering temperature for an undiluted system,
termed as Griffiths temperature $T_G$.
In doped manganites, quenched disorder that arises from the $A$ site disorder
is the reason for the random distribution of cations and the consequent random dilution.
In a similar fashion, the site disorder in double perovskites can lead to random
dilution of the $B$ lattice.
The negative curvature of 1/$\chi(T)$ in Fig. \ref{fig_chi} is observed to diminish
in magnitude as magnetic field increases from 20~Oe to 40~kOe.
This trend of 1/$\chi(T)$ conforms to the general features of a GP
phase where inverse susceptibility deviates from Curie-Weiss (CW) description as
$T \rightarrow T_c$ and is suppressed at higher magnetic field.
In Fig. \ref{fig_chi} (a), we identify $T_G$ ($\sim$ 164~K) as the characteristic
Griffiths temperature where 1/$\chi(T)$ commences to deviate from CW description.
The low-field magnetic susceptibility in the Griffiths phase follows
the power law behaviour, $\chi^{-1}(T) \propto (T - T^{R}_C)^{1-\lambda}$,
where $\lambda$ is the magnetic susceptibility exponent whose value lies in the
range 0 $< \lambda < $ 1.
The magnetic susceptibility of Tb$_2$NiMnO$_6$ at $H$ = 20~Oe was analyzed
using the power law with great care to avoid an
incorrect estimation of $T^R_C$ which can lead to erroneous values of $\lambda$.
Recent literature that offers a better protocol to perform the analysis
was adopted to estimate $T^R_C$ and $\lambda$ in
the purely paramagnetic region
\cite{zhoujpc_115_1535_2011,pramanikprb_81_024431_2010}.
Following the procedure reported in literature
\cite{zhoujpc_115_1535_2011,jiangepl_84_47009_2008}
we first estimated the value of $T^R_C$ in the purely paramagnetic
region above $T_G$.
A value of 56~K for $T^R_C$ was estimated in this way
which can be compared with that obtained by Zhou {\it et al.}.
\cite{zhoujpc_115_1535_2011}
Using this value of $T^R_C$, fitting was performed in the GP
regime to obtain a value of 0.88 for $\lambda$.
The result of the analysis is presented in Fig. \ref{fig_chi}(b) plotted as log(1/$\chi$)
versus $t_m$ = ($T - T^R_C$)/$T^R_C$.
In the high temperature regime, we obtained $\lambda$ = 0.08,
signifying that the system has entered a completely paramagnetic phase.
The Curie-Weiss analysis of the data above 200~K resulted in
effective moment value of $\mu_{eff}$ = 14.1~$\mu_{B}$.
The spin-only moments of Ni and Mn assuming the combination of Mn$^{4+}$ ($3d^3$, $S$ = 3/2)
and Ni$^{2+}$ ($3d^8$, $S$ = 1) yields $\mu_{eff}$ = 4.79 $\mu_B$ and the combination of Mn$^{3+}$ ($3d^4$, $S$ = 2)
and Ni$^{3+}$ ($3d^7$, $S$ = 3/2) results in $\mu_{eff}$ = 6.24 $\mu_B$.
%
\begin{figure}[!h]
\centering
\includegraphics[scale=0.55]{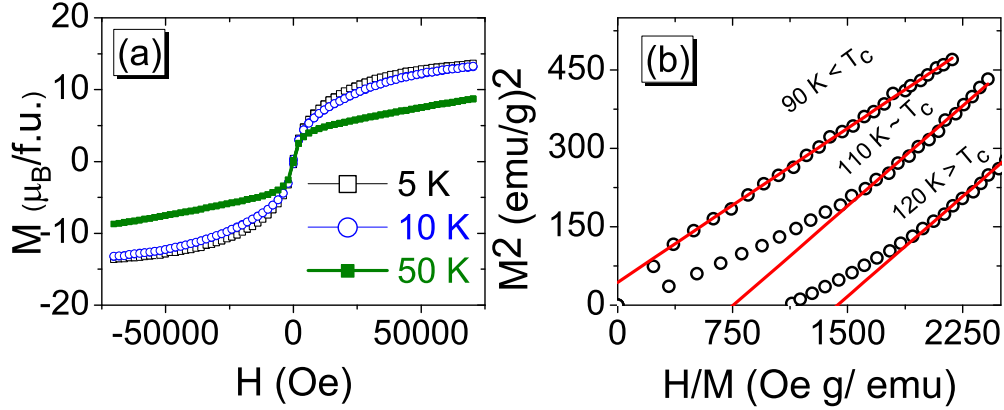}
\caption{(a) Shows the magnetization plots at 5, 10 and 50~K.
(b) The Arrott plots ($M^2$ versus $H/M$) at different
temperatures below and above $T_c$. The straight lines are the linear-fits to
the high-field region in the plot. No spontaneous magnetization is observed in the
temperature range near $T_c$ supporting the Griffiths phase-like characteristics.}
\label{fig_mag2}
\end{figure}
%
Assuming the rare earth paramagnetic moment to be 9.7~$\mu_B$
\cite{boothmrb_44_1559_2009}, the calculated
paramagnetic effective moment is $\approx$ 15.07~$\mu_B$
for the Mn$^{3+}$/Ni$^{3+}$ combination and 14.53~$\mu_B$ for the
Mn$^{4+}$/Ni$^{2+}$ combination.
A comparison with the experimentally observed value for
effective moment suggests that the cationic combination in
Tb$_2$NiMnO$_6$ is Mn$^{4+}$/Ni$^{2+}$.
The 1/$\chi(T)$ plot at an applied magnetic field of 20~Oe and the
corresponding CW fit are presented in Fig. \ref{fig_chi} (a).
\\
The isothermal magnetization plots of TNMO at 5, 10 and 50~K are
illustrated in Fig. \ref{fig_mag2} (a).
The qualitative nature of the curves are
different from the metamagnetic features observed at
low temperatures by Troyanchuk {\it et. al.,} in TbNi$_{0.5}$Mn$_{0.5}$O$_3$.
\cite{troyanchukmrb_32_67_1997}
The saturation magnetization $M_{sat}$ = 13.5 $\mu_{B}$ at the
highest applied field of 70~kOe is lower than
the theoretical value of 23.2 $\mu_B$ which is calculated as $2gJ + 5.2 \mu_B$
where, $g$ is the Land\'{e} g-factor, $J$ is the total angular momentum
and 5.2 $\mu_B$ is the effective Ni-Mn spin-only moment value.
\cite{boothmrb_44_1559_2009}
In La$_2$NiMnO$_6$, the saturation magnetization at highest applied
field is reported as 4.57~$\mu_B$ which is close to the theoretical value
of 5~$\mu_B$
\cite{goodenoughpr_124_373_1961}
but the value of $M_{sat}$ has been observed to vary with the preparation method.
\cite{dassprb_68_064415_2003}
%
\begin{figure}[!h]
\centering
\includegraphics[scale=0.55]{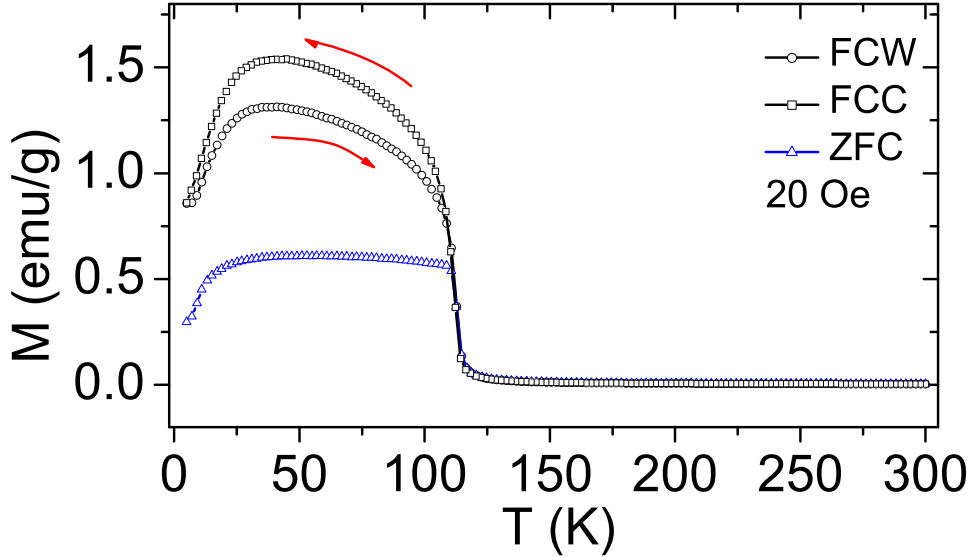}
\caption{The FCC and FCW cycles at 20~Oe showing a prominent hysteresis
of the first-order type resulting from the ferromagnetic clusters that form
below $T_c$.}
\label{fig_hyst}
\end{figure}
\\
One of the basic characteristics of Griffiths phase is the existence
of finite-size FM correlated spins but with no static long-range magnetization
which can be verified through the Arrott plots.
In Fig. \ref{fig_mag2} (b) we present the Arrott plots -- $M^2$ versus $H/M$ plots -- derived from
magnetization isotherms at 90, 110 and 120~K.
The solid lines in Fig. \ref{fig_mag2} (b) are linear fits to the high-field region
extrapolated to the axes so as to obtain the spontaneous magnetization.
We observe no spontaneous magnetization in the temperature region
close to $T_c$.
%
%
The Arrott plots display positive slope, characteristic of second-order
phase transitions as prescribed by the Banerjee criterion.
\cite{banerjeepl_12_16_1964,miraprb_60_2998_1999}
Generally, the $M^2$ versus $H/M$ plots are linear for homogeneous ferromagnets
\cite{herzerpss_101_713_1980}
and a deviation from linearity is suggestive of inhomogeneous behaviour.
The presence of inhomogeneous magnetism in Tb$_2$NiMnO$_6$ is confirmed
in Fig. \ref{fig_hyst}, where FCW and FCC curves at an applied field of
20~Oe exhibit prominent hysteresis.
Similar hysteresis behaviour of the FCC and FCW
arms has been reported in magnetically phase-separated La$_{0.5}$Ca$_{0.5}$MnO$_3$.
\cite{freitasprb_65_104403_2002}
\\
The structural distortions resulting from the substitution of a smaller
rare earth Tb at the $R$ site can couple with the spin
system, the signature of which is discernible in Raman scattering experiments.
Hence, we performed Raman scattering experiments on Tb$_2$NiMnO$_6$.
The experimentally observed scattering intensity as a function
of Raman shift is plotted in Fig. \ref{fig_raman} (a)
for different temperatures.
In Fig. \ref{fig_raman} (b), the intensity at 298~K is plotted as a function
of Raman shift and along with deconvoluted peaks
assuming Lorenzian peak shapes.
Following earlier reports
\cite{guoapl_89_022509_2006,bulljssc_177_2323_2004}
which established the similarity of Raman spectra of double
perovskite La$_2$NiMnO$_6$ with orthorhombic LaMnO$_3$, we assign the respective
peaks at 645 cm$^{-1}$ and 490 cm$^{-1}$ to {\it stretching} and
{\it anti-stretching} vibrations of (Ni/Mn)O$_6$ octahedra.
\cite{bulljssc_177_2323_2004}
Lattice dynamical calculations attribute mixed character to the
490 cm$^{-1}$ mode which involves both anti-stretching and bending
vibrations whereas the 645 cm$^{-1}$ mode is purely a stretching mode.
\cite{ilievprb_75_104118_2007}
The Raman spectra of Tb$_2$NiMnO$_6$ is similar to that of
Pr$_2$NiMnO$_6$ which also crystallizes in monoclinic $P2_1/n$ symmetry.
\cite{truongjpcm_23_052202_2011}
The temperature dependence of phonon frequencies, $\omega(T)$, are
shown in Fig. \ref{fig_fwhm} (a) along with curve fits
assuming a standard
\cite{balkanskiprb_28_1928_1983}
anharmonic dependence of phonon modes,
$\omega_{anh}(T) = \omega_0 - C[1 + 2/(e^{\hbar\omega/k_BT} - 1)]$,
where, $\omega_0$ is the temperature-independent part of linewidth,
$C$ is a constant determined from curve-fit, $\hbar\omega$ is the phonon energy and  $k_B$ is
the Boltzmann constant.
$R$MnO$_3$ perovskites which are $A$ type antiferromagnets and ferromagnetic double
perovskites like La$_2$NiMnO$_6$ (LNMO) and La$_2$CoMnO$_6$ (LCMO) exhibit
magnetic-order-induced phonon renormalization.
\cite{laverdiereprb_73_214301_2006,granadoprb_60_11879_1999,ilievprb_75_104118_2007,ilievjap_106_023515_2009,truongprb_76_132413_2007}
These systems display mode softening and a deviation from the
$\omega_{anh}(T)$ dependence of phonon frequencies below $T_c$.
However, in $R$MnO$_{3}$ systems with an incommensurate
magnetic structure ($R$ = Eu, Tb or Y) no mode softening is observed.
\cite{laverdiereprb_73_214301_2006}
However, the data on Tb$_2$NiMnO$_6$ yields a reasonably good fit to
the standard expression for anharmonic dependence of phonons
showing that the predominant effects in the phonon spectrum are anharmonic.
The reduction in spin-phonon coupling is correlated to the size of the $R$ ion as
strength of spin-phonon coupling shows a decreasing trend
with reduction in rare earth size as reported in the case of Pr$_2$NiMnO$_6$.
\cite{truongjpcm_23_052202_2011}
In Fig. \ref{fig_fwhm} (b), the temperature variation of full-width-at-half-maximum
(FWHM) for pure {\it stretching} mode at 645 $cm^{-1}$
is plotted.
The linewidth of phonon is a measure of the phonon lifetime and is determined
by temperature-dependent scattering from lattice defects or phonons.
%
For Tb$_2$NiMnO$_6$, as clear from Fig. \ref{fig_fwhm} (b), we observe three different
regions in FWHM, with discontinuities at $T \sim$ 110~K and at 180~K.
Note that the discontinuities occur at the same value of temperature
where the deviation from perfect CW description occurs in the
$1/\chi(T)$ curves.
This indicates a close correlation between magnetism
and the lattice in Tb$_2$NiMnO$_6$.
\\
To summarize the results, we have refined the crystal structure of Tb$_2$NiMnO$_6$
in monoclinic $P21/n$ space group that allows for a Ni$^{2+}$/Mn$^{4+}$ cationic arrangement.
The estimated bond distances and the BVS values indicate a nearly-ordered
$B$ site structure.
The downturn of 1/$\chi(T)$ curves from Curie-Weiss behaviour which
decreases with the application of magnetic field
is reminiscent of Griffiths phase in disordered systems.
The presence of GP in Tb$_2$NiMnO$_6$ is further supported by the power-law
analysis of 1/$\chi(T)$ and the Arrott plots.
On a microscopic level, parameters like the magnetic correlation length
($\xi$) derived from small-angle neutron scattering experiments
can further illuminate about the Griffiths phase.
\cite{heprb_76_014401_2007}
For a conventional FM phase transition, $\xi (T)$ displays a gradual increase
from zero at high $T$, then diverges as  $T \rightarrow T_c$
whereas in the presence of spin correlations, it shows a sharp increase at
the temperature where the clusters emerge.
Electron spin resonance (ESR) studies also have been successfully
employed to study the GP phase in disordered manganites.
\cite{deisenhoferprl_95_257202_2005}
At this point, it is instructive to compare the case of La$_2$NiMnO$_6$, in which the
short range ferromagnetic correlations above $T_c$ have been studied
through x ray magnetic circular dichroism,
\cite{guoprb_79_172402_2009}
critical behaviour,
\cite{luojpcm_20_465211_2008critical}
and ESR.
\cite{zhouapl_91_172505_2007}
Contrary to the case of Tb$_2$NiMnO$_6$, a positive deviation of $1/\chi(T)$ from
CW description was observed.
However, it has been demonstrated that the degree of antisite
disorder is a function of the kind of cation
at the $R$ site and can be influenced through doping.
\cite{garciahernandezprl_86_2443_2001}
%
%
Analysis of the Raman scattering data showed that, in contrast
to the behaviour seen in other DP like La$_2$NiMnO$_6$,
magnetic-order-induced mode softening is not observed in Tb$_2$NiMnO$_6$.
Normally, FWHM of Raman linewidth decreases with temperature as indicated
by the red dashed line in Fig. \ref{fig_fwhm} (b).
However, we observe a marked deviation from
linear behaviour below about 180~K, extending till $T_c$.
In this region, FWHM shows a plateau-like region (black solid line in Fig. \ref{fig_fwhm} (b)).
The anomalous plateau-like behaviour of TNMO  is suggestive of increase in phonon lifetime
which originates from the disorder in the system that strongly couples the magnetic and lattice degrees of freedom.
The antisite disorder leads to additional antiferromagnetic exchange
interactions between Ni -- Ni and Mn -- Mn.
It is significant to note that the range of temperature where
deviation from linearity in FWHM 
occurs (Fig. \ref{fig_fwhm} (b)) coincides with the range over
which $1/\chi(T)$ deviates from CW description (Fig. \ref{fig_mag2}).
Our study projects that Raman scattering can be used as a
tool for the characterization of materials exhibiting
magneto-lattice coupling effects.
The temperature variation of FWHM  for other frequencies (presented in Fig. \ref{fig_fwhm} (b))
do not show slope changes which points towards strong spin-phonon coupling
present in the case of symmetric modes like the {\it stretching} mode.
The role of Tb-magnetism in Tb$_2$NiMnO$_6$ is not significant since ESR studies
have shown that the Ni--Mn sublattice interacts weakly with the rare earth.
\cite{boothmrb_44_1559_2009}
However, the small cationic radius of Tb has a marked impact on structural distortions
and contribute to the spin-lattice coupling leading to stronger
magnetocapacitive effects.
\cite{rogadoam_17_2225_2005}
This motivates us to investigate further the magneto-dielectric
properties of this compound, in detail with emphasis on
multiferroism in double perovskites.
\section{Conclusions}
%
We observe Griffiths phase-like features in the magnetic properties of
double perovskite Tb$_2$NiMnO$_6$ where the smaller rare earth
size of Tb influences the spin-phonon coupling.
The value estimated for the inverse susceptibility exponent $\lambda$ along with the
absence of spontaneous magnetization observed through Arrott plots
testify the inhomogeneous nature of magnetism in this material
and confirm the Griffiths phase-like features.
The observed features arise from the site disorder at the $B$ site
supplemented by the lattice distortions brought
about by the smaller radius of Tb.
No mode softening of phonon frequencies is observed in Tb$_2$NiMnO$_6$
which is qualitatively different from the
observation in other double perovskites with a non-magnetic rare earth.
Thus, we conclude that the magnetic and lattice properties of $R_2BB'$O$_6$
where $R$ is a magnetic rare earth are different from those of the non-magnetic counterparts.
\section*{Acknowledgments}
SE wishes to thank Department of Science and Technology, India for financial support through project grants.\\
%

\end{document}